\documentclass[useAMS,usenatbib]{mnras}
\usepackage{graphicx} 
\usepackage{amsmath} 
\usepackage{amssymb} 
\usepackage{multicol} 
\usepackage{bm} 
\usepackage{pdflscape} 
\usepackage{float}
\usepackage{gensymb}
\usepackage{xspace}

\newcommand{\kms}{$\mbox{km s}^{-1}$\xspace}
\newcommand{\halpha}{\mbox{H$\alpha$}\xspace}
\newcommand{\lambdaf}{\,$\lambda$}%
\newcommand{\heI}{\mbox{\ion{He}{i}}\xspace}
\newcommand{\heII}{\mbox{\ion{He}{ii}}\xspace}
\newcommand{\CIII}{\mbox{\ion{C}{iii}}\xspace}
\newcommand{\OII}{\mbox{\ion{O}{ii}}\xspace}
\newcommand{\MgII}{\mbox{\ion{Mg}{ii}}\xspace}
\newcommand{\SiIII}{\mbox{\ion{Si}{iii}}\xspace}
\newcommand{\SiIV}{\mbox{\ion{Si}{iv}}\xspace}
\newcommand{\asource}{\mbox{A0538--66}\xspace}
\newcommand{\vsini}{$v \sin i$\xspace}
\newcommand{\teff}{$T_{\rm eff}$\xspace}
\newcommand{\logg}{$\log{g}$\xspace}

\newcommand{\msun}{\mbox{M$_\odot$}\xspace}
\newcommand{\rsun}{\mbox{R$_\odot$}\xspace}
\newcommand{\MyChi}{\raisebox{0.35ex}{\( \chi \)}}%

\usepackage[T1]{fontenc}
\usepackage{ae,aecompl}

\def\aap{Astronomy \& Astrophysics}

\title[Be/X-ray binary \asource]{Orbital and Superorbital Monitoring of the Be/X-ray binary \asource: constraints on the system parameters \thanks{Based on observations made with the Southern African Large Telescope (SALT).}}
\author[A. F. Rajoelimanana et al.]{A. F. Rajoelimanana$^{1}$\thanks{E-mail: andry@saao.ac.za}, P. A. Charles$^{2,3,4}$, P. J. Meintjes$^{1}$, L. J. Townsend$^{4}$, \newauthor M.P.E. Schurch$^{4}$ and A. Udalski$^{5}$\\\\\\
$^{1}$Department of Physics, University of the Free State, PO Box 339, Bloemfontein 9300, South Africa \\
$^{2}$School of Physics and Astronomy, Southampton University, Southampton SO17 1BJ \\
$^{3}$Department of Astrophysics, University of Oxford, Keble Road, Oxford OX1 3RH, UK \\
$^{4}$Astrophysics, Cosmology and Gravity Centre, Department of Astronomy, University of Cape Town, Rondebosch 7701, South Africa \\
$^{5}$Warsaw University Observatory, Aleje Ujazdowskie 4, 00-478 Warsaw, Poland}
\date{30 September 2016, accepted for publication in MNRAS}
\pubyear{2016}
\begin{document}
\label{firstpage}
\pagerange{\pageref{firstpage}--\pageref{lastpage}}
\maketitle
\begin{abstract}
We combine the decade long photometry of the Be/X-ray binary system \asource provided by the MACHO and OGLE~IV projects with high resolution SALT spectroscopy to provide detailed constraints on the orbital parameters and system properties.  The $\sim$420d superorbital modulation is present throughout, but has reduced in amplitude in recent years.  The well-defined 16.6409d orbital outbursts, which were a strong function of superorbital phase in the MACHO data (not occurring at all at superorbital maximum), are present throughout the OGLE~IV coverage.  However, their amplitude reduces during superorbital maximum.  We have refined the orbital period and ephemeris of the optical outburst based on $\sim$ 25 yrs light curves to  HJD~=~2455674.48$\pm$0.03 + n*16.6409$\pm$0.0003d.  Our SALT spectra reveal a B1~III star with \vsini of~285~\kms from which we have derived an orbital radial velocity curve which confirms the high eccentricity of $e$ = 0.72$\pm$0.14.  Furthermore, the mass function indicates that, unless the neutron star far exceeds the canonical 1.44 \msun, the donor must be significantly undermassive for its spectral type.  We discuss the implications of the geometry and our derived orbital solution on the observed behaviour of the system.

\end{abstract}

\section{Introduction}

The recurrent X-ray transient \asource was first discovered in 1977 by \citet{White1978} when two X-ray outbursts were detected with the {\it Ariel V} satellite. Subsequently, several outbursts from the source which recur with a period of 16.7 d were observed with the {\it HEAO-1} modulator collimator \citep{Johnston1979, Johnston1980, Skinner1980}. The improved {\it HEAO-1} position enabled \citet{Johnston1980} to identify its optical counterpart with a bright (V$\sim$15.7) Be star. The radial velocity of its absorption lines confirms its physical association with the Large Magellanic Cloud (LMC), and hence a well-established distance of 50 kpc \citep{Alves2004}. At this distance the peak luminosity of the X-ray outbursts were estimated to be $\sim$ $\mathrm{10^{39} erg~s^{-1}}$. That makes \asource one of the most luminous X-ray sources known, and still the highest in the Be/X-ray binary (BeX) class \citep{Liu2005}. In one of the X-ray outbursts observed with the Einstein Observatory, \citet{Skinner1982} found an X-ray pulsation with a period of 69 ms which indicates the presence of a rapidly rotating NS in the system.  However, for a distance of 50 kpc the observed X-ray luminosity is then substantially super-Eddington for a 1.44 $\mathrm{M_{\odot}}$ neutron star (NS).

\citet{Skinner1981} derived an improved orbital period of 16.6515 d using archival plates taken with the UK Schmidt telescope. They also reported that the X-ray outbursts were accompanied by an optical brightening of 2 mag. \citet{Densham1983} observed four consecutive outbursts in late 1981, from which they found that the source can reach as bright as V$\sim$12, and that the optical brightening was accompanied by the sudden appearance of very strong \heII $\mathrm{\lambda4686}$ emission. Since late 1983, the scale of these periodic outbursts has reduced considerably, with only much weaker outbursts (few tenths of a magnitude) being seen in both optical and X-ray observations.

Using optical and UV spectroscopy, \citet{Charles1983} classified the optical counterpart of \asource as a B2~III--V star. However, \citet{Hutchings1985} and \citet{Negueruela2002} suggested slightly earlier spectral types of B1 and B0.5~III, respectively, based on optical spectroscopic data taken during quiescence.

The existence of long-term optical monitoring of the LMC by gravitational micro-lensing projects, such as the MACHO, has revolutionized our understanding of the long-term behaviour of \asource. \citet{Alcock2001} analysed the first $\sim$5 years MACHO light curve and found that the source displays large amplitude superorbital variability on a timescale of 420.8d.
This was suggested to be a result of the formation and dissipation of the equatorial disc around the donor star. Analysis of the 70 years of archival Harvard and Schmidt plates by \citet{McGowan2003} also showed evidence of this long-term modulation. Within this 420.8d superorbital variation, the 16.6515d orbital outbursts were seen, but they were confined to only occur during optical minimum, never during the optical maximum (when the source remained quiescent). At optical maximum \asource is bluer, but remarkably when the disc forms it masks part of the hotter Be star which makes the source appear redder and fainter. This behaviour is typical for an equatorial disc viewed at high inclination angle \citep[for details, see][]{Rajoelimanana2011}. Based on the brightness changes between the optical maxima and minima, \citet{McGowan2003} suggested a lower limit for the inclination of the Be star of $i=74.9\degr$.

Given its relative brightness and now well-established orbital period, there have been attempts to obtain spectroscopic radial velocity curves in order to constrain the system's kinematic properties \citep{Corbet1985, Hutchings1985}.  However, with a likely rather high eccentricity proposed by \citet{Charles1983} (on the basis of the orbital optical/X-ray light curves) and a neutron star compact object, significant velocity changes were only expected close to periastron.  Unfortunately, the radial velocity curves of \citet{Corbet1985} and \citet{Hutchings1985} are inconsistent, for which \citet{Smale1989} suggested that the spectra from that time may have been contaminated by residual emission components.

This paper presents an analysis of the complete MACHO \citep{Alcock1996} and OGLE~IV \citep{Udalski2015} optical light curves of \asource as well as the results of optical spectroscopy obtained from the Southern African Large Telescope (SALT) and the SAAO 1.9 m telescope. The spectra are of sufficient resolution and S/N to obtain a new radial velocity curve, which is similar to that of \citet{Hutchings1985}.

\section{Observations and data reductions}

\subsection{Optical light curve}
\label{optical_lightcurve}

\begin{figure}
\scalebox{0.5}{\includegraphics{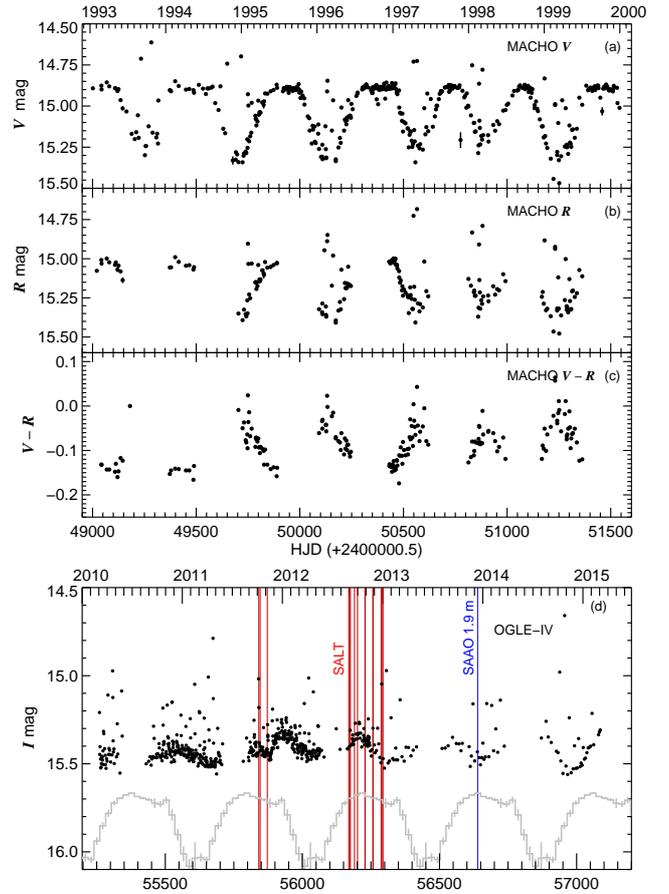}}
\caption{MACHO $V$ (a) and $R$ (b) light curves. (c) MACHO colour variation, (d) OGLE~IV light curves together with the mean MACHO $V$ light curve (in grey) projected forward using the ephemeris of \citet{Alcock2001}. The red and blue vertical lines refer to the times of the SALT and SAAO 1.9 m telescope observations respectively. See Fig.~\ref{plot_fithalpha} for more detail regarding the timing of the SALT spectra within the overall OGLE~IV 
light-curve}
\label{lcall} 
\end{figure}

The MACHO observations were taken between 1992 and 2000 using the 1.27 m telescope at Mount Stromlo Observatory, Australia \citep{Alcock1996}. Observations were performed in two pass-bands, a blue band ($\sim$~4500~--~6300 \AA, slightly shorter effective wavelength than Johnson V) band and a red band ($\sim$~6300~--~7600 \AA, a slightly longer effective wavelength than Johnson R). The complete MACHO light curves are provided in instrumental magnitudes and are publicly available at the MACHO website\footnote{http://macho.anu.edu.au}. We transformed the MACHO blue and red data to the standard Kron-Cousins $V$ and $R$ systems using a calibration procedure as described in \citet{Alcock1999}. \asource is identified with the object 61.9045.32 in the MACHO catalogue. 
The field of \asource also lies within the field regularly monitored by the OGLE~IV survey \citep{Udalski2015} in which the source is designated as the OGLE object LMC~518.26~21298. Its real-time OGLE~IV {\it I}--band light curve is available at the X-ray variables OGLE monitoring (XROM) website\footnote{http://ogle.astrouw.edu.pl/ogle4/xrom/xrom.html} \citep{Udalski2008}. Unfortunately, our source is not in the OGLE~II and OGLE~III survey regions.

\begin{figure}
\scalebox{0.5}{\includegraphics{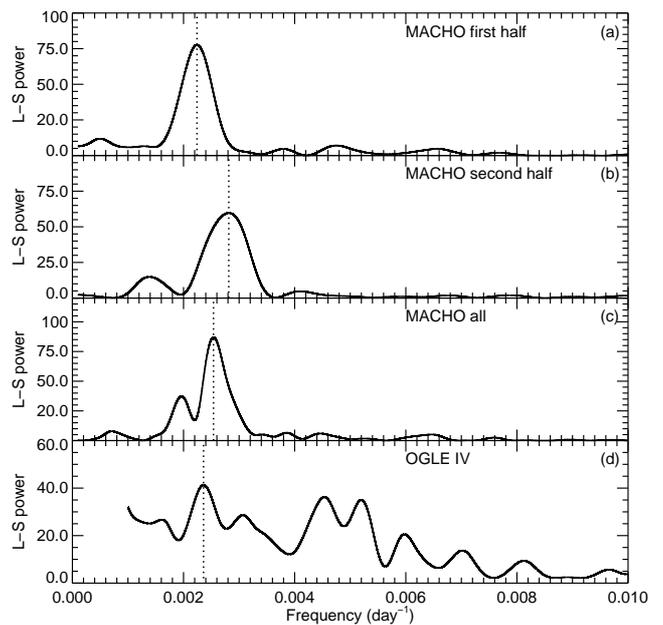}}
\caption{Lomb-Scagle periodogram of the MACHO $V$ data showing peaks at 445.1d, 352.7d and 391.4d for the first half (a), second half(b) and complete light curve (c) respectively. (d) Power spectrum of the OGLE~IV data showing a peak at 423.74d}
\label{periodlong} 
\end{figure}

Fig.~\ref{lcall} shows the complete MACHO ($V$ and $R$) and OGLE~IV light curves as well as the colour variation (MACHO $V$-$R$) of \asource. The MACHO light curve clearly displays a long-term variation of very large amplitude ($\sim$0.5 mag). However, the 420.8d superorbital period reported by \citet{Alcock2001} in the early 5 years of the MACHO light curve appeared to have slightly shortened in the second half of the data. We have removed the periodic 16.64 d optical outburst signal by taking out all data points with phase between 0.95 and 1.15 and performed a Lomb-Scargle periodogram analysis to search for periodicities. The power spectrum of the complete MACHO $V$ data shows a peak at 393.85~$\pm$~0.14d (Fig.~\ref{periodlong}) which is slightly shorter than the previously reported 420.8d period obtained from analysing all the MACHO data, i.e. including the orbital outburst points.

\begin{figure}
\scalebox{0.5}{\includegraphics{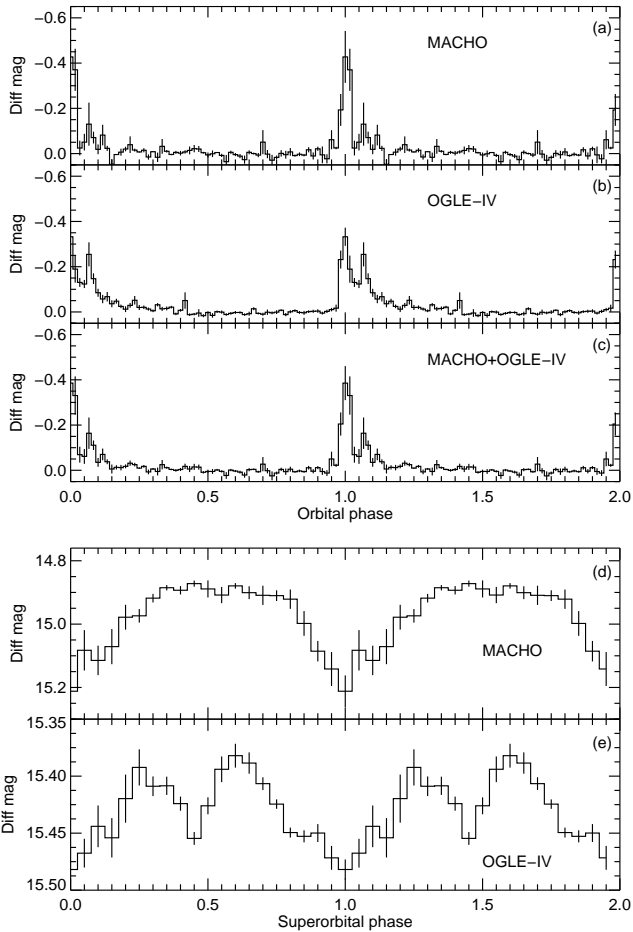}}
\caption{MACHO (a), OGLE~IV (b) and combined (c) light curves folded on the orbital period of 16.6409d. (d) and (e): MACHO and OGLE IV light curves folded on the 393.85d and 423.74d superorbital periods, respectively.}
\label{fold_all} 
\end{figure}

To further explore the variation in the timescale of the superorbital modulation, the complete MACHO data were divided into two parts. The power spectrum of the first and second half shows a peak at 446.0~$\pm$~0.2d and 355.4~$\pm$~0.29d respectively (Fig.~\ref{periodlong}). The OGLE~IV also shows a long-term variation but now of a much smaller amplitude, although still with a period consistent with that in the MACHO data. When all bright points related to the 16.64d outburst were removed (as in the MACHO data) the power spectrum shows its highest peak at $P_{\rm sup}$ = 423.74~$\pm$0.42d and some additional peaks around 200d (see Fig.~\ref{periodlong} ). We have compared this behaviour to the original superorbital period by projecting the 420.8d light curves and ephemeris by \citet{Alcock2001} into the OGLE~IV data (Fig.~\ref{lcall}).  The optical maxima in the MACHO data are flat (i.e. no optical outbursts were seen) and last $\sim$200 d, meaning that we are seeing the completely disc-less B star. However in the OGLE IV phase, the optical maxima are not flat and the orbital outbursts now occur throughout the OGLE superorbital cycle, implying that during the OGLE~IV optical maxima the equatorial disc was not completely dissipated. The MACHO and OGLE IV light curves folded on the superorbital period are shown in Fig~\ref{fold_all}.

The colour variation ($V - R$) of \asource shows that the source is bluer at optical maxima and becomes redder as the source fades. This behaviour is opposite to that seen in most other BeX where the source is redder when it is brighter. However, during outbursts on the orbital period associated with periastron passage of the neutron star, the optical brightening is accompanied by a slight increase in reddening (i.e. the source gets redder).

The amplitude of the outbursts in the MACHO data clearly varies throughout the long-term 420.8d cycle, being very strong ($\sim$0.7 mag) at optical minima, decreasing in amplitude as the source brightens and completely disappearing at optical maxima. Similar behaviour is seen in the OGLE~IV data; however, there are clearly orbital outbursts occurring even when the source is at maximum in the long-term cycle.

\begin{figure}
\scalebox{0.5}{\includegraphics{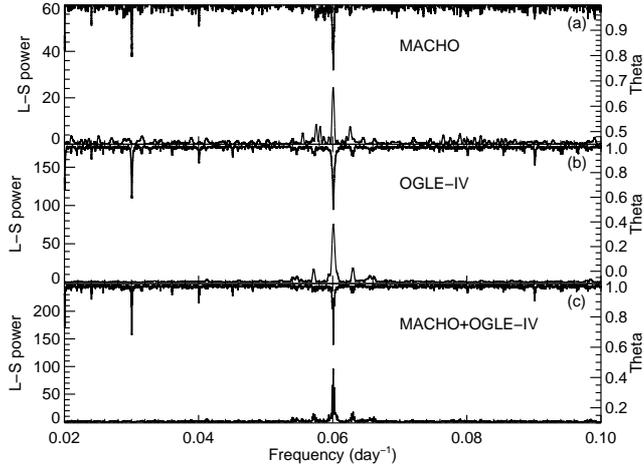}}
\caption{{\it First panel}: PDM statistic ({\it upper plot}) and L-S periodogram ({\it lower plot}) for the MACHO data. {\it Second panel}: similar to the first panel but for OGLE~IV data. {\it Bottom panel}: PDM statistic and L-S periodogram for the combined light curves showing the peak at 16.6409~$\pm$~0.0003 d.}
\label{periodshort} 
\end{figure}

In order to refine the orbital period, we removed the long-term trend in the light curves. To do this, the data were split into several segments, the periodic outburst events were first removed by sigma-clipping the bright points for each segment, from which we fitted and subtracted a low-order polynomial. Lomb-Scargle periodogram and phase dispersion minimization were performed on both individual and combined datasets. The highest peak in the periodogram of the detrended OGLE~IV data occurs at 16.6398~$\pm$~0.0026d) which is slightly shorter than that found in the MACHO data (P=16.6434~$\pm$~0.0024d), but consistent within the uncertainties. For the combined datasets, the highest peak occurs at 16.6409~$\pm$~0.0003d (see Fig.~\ref{periodshort}). We estimate an ephemeris for maximum light as HJD$_{\mathrm{max}}$ = 2455674.48~$\pm$~0.03 + n*16.6409~$\pm$~0.0003d. Throughout this paper, we use this new ephemeris so that phase~0 is the phase of maximum brightness. In Fig.~\ref{fold_all}, we have folded the light curves into 60 phase bins using the new ephemeris so as to better display the structure in the orbital light curve during outbursts. The orbital profiles clearly show two peaks during optical outbursts which indicates a misalignment between the Be equatorial disc and the NS orbital plane \citep{Coe2008} and which will be discussed further in section~\ref{bephase}

\subsection{The spectroscopic data}

We obtained spectroscopic observations for \asource with the Robert Stobie Spectrograph \citep[RSS:][]{Burgh2003,Kobulnicky2003} on the 10 m Southern African Large Telescope (SALT) \citep{Buckley2006} in Sutherland, South Africa. The observations were carried out in long-slit spectroscopy mode between Aug 2011 and Jan 2013, using two volume phase holographic (VPH) gratings (PG0900 and PG3000) on each observing night. Our detailed setups were (see also Table~\ref{tab:highres_log}):

\begin{itemize}
\item{{\it Low-resolution} spectra: These were obtained by combining spectra from two grating angles (12.5, 20.0\degr ) of {\sc PG0900} (900 lines mm$^{-1}$), yielding coverage from $\sim$3100 -- 9000 \AA\ at a dispersion of $\sim$ 1.0 \AA~pixel$^{-1}$ and a spectral resolution of  190 \kms ($\sim$ 4.0 \AA) at \halpha.
}
\item{{\it High-resolution blue} spectra: The 3000 lines mm$^{-1}$ grating ({\sc PG3000}) was used with a slit width of 0.6\arcsec\ to provide a spectral resolution of  50 \kms (0.73 \AA).
}
\end{itemize}

The data from each 2048$\times$4096 CCD detector were pre-reduced with the {\sc iraf} package {\sc pysalt} \citep{Crawford2010} which includes overscan subtraction, gain and cross-talk corrections and mosaicking. The cosmic ray events were cleaned using the routine {\sc l.a.cosmic} \citep{Van_dokkum2001}. We applied a wavelength solution using arc lamp spectra ; the resulting wavelength accuracy was about 0.025~\AA. Flat-field correction and background subtraction were performed using the standard {\sc iraf} procedures. One-dimensional spectra were then extracted using the {\sc iraf} task {\sc apall}. Owing to the moving pupil design of SALT the effective area of the telescope is changing during exposures. This means that absolute flux calibration is not possible. However, a relative flux correction to recover the spectral shape was applied to the extracted spectra using the observed spectrophotometric standard. We then combine multiple exposures obtained at the same instrument setting during the same visit.

Additional high-resolution spectra were obtained at phase~0.993 on 2013 Dec 13 (MJD = 56639.0626) using the grating spectrograph on the 1.9 m Radcliffe telescope of the South African Astronomical Observatory (SAAO). The detector was a 266$\times$1798 {\sc sit}e {\sc ccd} used in 2$\times$1 binning mode. A 1200 lines mm$^{-1}$ reflection grating blazed at 4600~\AA\ (Grating~\#4) was used in the spectral range of 4100~--~5000~\AA. The observations were carried out at an airmass of $\sim$1.25 with a slit width of 1.2\arcsec, yielding a spectral resolution of 65~\kms ($\sim$ 1 \AA). The data were reduced using standard {\sc iraf} packages.

\begin{table}
\centering
\caption[]{Log of the SALT spectroscopic observations.}
\label{tab:highres_log}
\scriptsize
\begin{tabular}{ccccccc}
\hline
\hline
\multicolumn{1}{c}{Date} &
\multicolumn{1}{c}{MJD} &
\multicolumn{2}{c}{Grating} &
\multicolumn{1}{c}{Exp$^{\star}$} &
\multicolumn{2}{c}{Phase} \\

\multicolumn{1}{c}{of Obs} &
\multicolumn{1}{c}{} &
\multicolumn{2}{l}{name~angle($\degr$)} &
\multicolumn{1}{c}{(s)} &
\multicolumn{1}{c}{orb$^{\dagger}$} &
\multicolumn{1}{c}{sup$^{\ddagger}$} \\

\hline

2011-10-07    &   55841.020    &    PG0900    &      12.5    &   120    &     0.038    &     0.253  \\
         --           &   55841.027    &    PG0900    &      20.0    &   120    &     0.038    &     0.253  \\
         --           &   55841.034    &    PG3000    &      44.0    &   700    &     0.039    &     0.253  \\
2011-10-12    &   55846.013    &    PG0900    &      12.5    &   120    &     0.338    &     0.264  \\
         --           &   55846.019    &    PG0900    &      20.0    &   120    &     0.338    &     0.264  \\
         --           &   55846.025    &    PG3000    &      44.0    &   700    &     0.339    &     0.264  \\
2011-11-07    &   55872.028    &    PG0900    &      12.5    &    60    &     0.901    &     0.326  \\
         --           &   55872.034    &    PG0900    &      20.0    &    80    &     0.902    &     0.326  \\
         --           &   55872.038    &    PG3000    &      44.0    &   700    &     0.902    &     0.326  \\
2012-08-30    &   56169.121    &    PG0900    &      12.5    &   375    &     0.754    &     0.032  \\
         --           &   56169.132    &    PG0900    &      20.0    &   360    &     0.755    &     0.032  \\
         --           &   56169.143    &    PG3000    &      41.0    &   840    &     0.756    &     0.032  \\
2012-09-02    &   56172.133    &    PG0900    &      12.5    &   375    &     0.935    &     0.039  \\
         --           &   56172.141    &    PG0900    &      20.0    &   360    &     0.936    &     0.039  \\
         --           &   56172.151    &    PG3000    &      41.0    &   840    &     0.937    &     0.039  \\
2012-09-05    &   56175.119    &    PG0900    &      12.5    &   375    &     0.115    &     0.047  \\
         --           &   56175.126    &    PG0900    &      20.0    &   360    &     0.115    &     0.047  \\
         --           &   56175.136    &    PG3000    &      41.0    &   840    &     0.116    &     0.047  \\
2012-09-20    &   56190.085    &    PG0900    &      12.5    &   375    &     0.014    &     0.082  \\
         --           &   56190.096    &    PG0900    &      20.0    &   360    &     0.015    &     0.082  \\
         --           &   56190.109    &    PG3000    &      41.0    &   840    &     0.016    &     0.082  \\
2012-10-01    &   56201.037    &    PG0900    &      12.5    &   300    &     0.672    &     0.108  \\
         --           &   56201.045    &    PG0900    &      20.0    &   270    &     0.673    &     0.108  \\
         --           &   56201.057    &    PG3000    &      41.0    &   800    &     0.674    &     0.108  \\
2012-10-27    &   56227.967    &    PG0900    &      12.5    &   450    &     0.291    &     0.172  \\
         --           &   56227.974    &    PG0900    &      20.0    &   450    &     0.291    &     0.172  \\
         --           &   56227.985    &    PG3000    &      41.0    &  1200    &     0.292    &     0.172  \\
2012-10-28    &   56228.965    &    PG0900    &      12.5    &   450    &     0.351    &     0.174  \\
         --           &   56228.975    &    PG0900    &      20.0    &   360    &     0.351    &     0.174  \\
2012-11-25    &   56256.877    &    PG0900    &      12.5    &   400    &     0.028    &     0.241  \\
         --           &   56256.887    &    PG0900    &      20.0    &   340    &     0.029    &     0.241  \\
         --           &   56256.894    &    PG3000    &      41.0    &  1000    &     0.029    &     0.241  \\
2012-11-26    &   56257.917    &    PG0900    &      12.5    &   400    &     0.090    &     0.243  \\
         --           &   56257.926    &    PG0900    &      20.0    &   340    &     0.091    &     0.243  \\
         --           &   56257.935    &    PG3000    &      41.0    &  1000    &     0.092    &     0.243  \\
2012-12-25    &   56286.878    &    PG0900    &      12.5    &   400    &     0.831    &     0.312  \\
         --           &   56286.885    &    PG0900    &      20.0    &   340    &     0.831    &     0.312  \\
         --           &   56286.893    &    PG3000    &      41.0    &  1000    &     0.832    &     0.312  \\
2012-12-28    &   56289.944    &    PG0900    &      12.5    &   400    &     0.015    &     0.319  \\
         --           &   56289.955    &    PG0900    &      20.0    &   340    &     0.016    &     0.319  \\
         --           &   56289.964    &    PG3000    &      41.0    &  1000    &     0.016    &     0.319  \\
2013-01-01    &   56293.820    &    PG0900    &      12.5    &   800    &     0.248    &     0.329  \\
         --           &   56293.832    &    PG0900    &      20.0    &   680    &     0.249    &     0.329  \\
         --           &   56293.845    &    PG3000    &      41.0    &  1200    &     0.249    &     0.329  \\

\hline

\multicolumn{7}{l}{$^{\star}$ Total exposure time.}\\
\multicolumn{7}{l}{$^{\dagger}$ Orbital phases of the observations using our orbital ephemeris.}\\
\multicolumn{7}{l}{$^{\ddagger}$ Superorbital phases using the ephemeris of \citet{Alcock2001}}\\

\end{tabular}

\end{table}

\section{The optical spectrum}

\subsection{Broad-band spectra}

\begin{figure*}
\scalebox{0.5}{\includegraphics{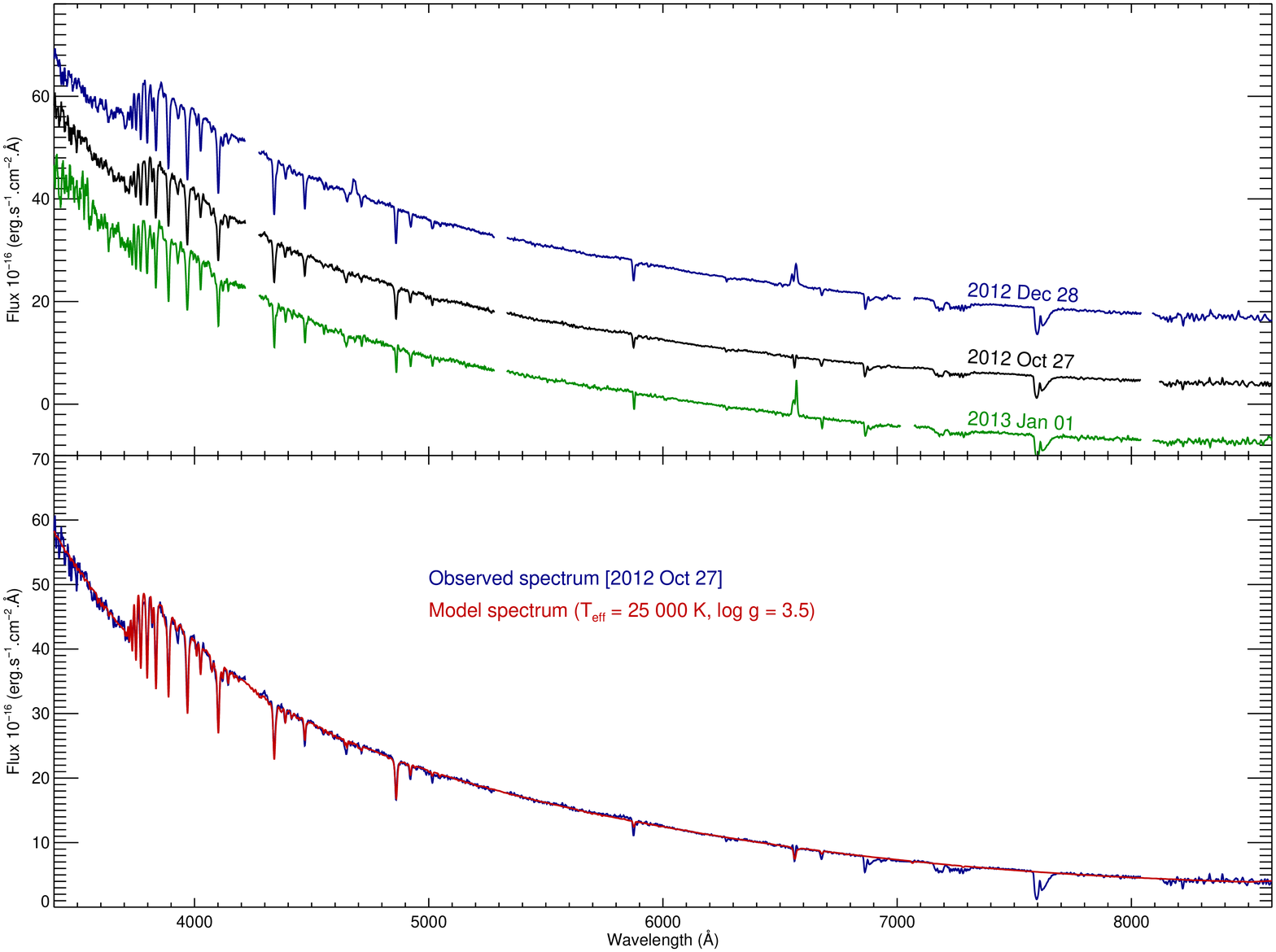}}
\caption{{\it Top:} Broad-band spectra of \asource taken during outburst (2012 Dec 28, blue), near optical maximum (2012 Aug 27, black) and during the Be phase (2013 Jan 01, green). Note the presence of the \heII~$\lambda$4686 in the spectrum obtained during outburst. {\it Bottom:} Spectrum of \asource taken on the night of 2012 Aug 27 (near optical maximum) compared to the best fit model with \teff = 25000 K and \logg = 3.5. The observed spectra were de-reddened using an extinction of $E$($B$ -- $V$) = 0.12. The gaps in the spectra are CCD gaps.}
\label{plot_broad} 
\end{figure*}

Spectra of Be stars are characterized by Balmer (and sometimes \heI)  emission lines and infrared excess originating from the circumstellar disc of gas surrounding the rapidly rotating star. We combine the two spectra obtained from two different settings of PG0900 in order to obtain one wide-wavelength spectra covering $\sim$3100--9000 \AA. The spectra were then de-reddened using an extinction of $E$($B$ -- $V$) = 0.12 \citep{Dutra2001}. Fig.~\ref{plot_broad} (top panel) shows an example of combined spectra obtained near optical maximum on (2012 Oct 27), during the Be phase (2013 Jan 01) and at outburst (2012 Dec 28). The spectrum obtained near optical maximum is dominated by the photospheric spectrum of the underlying B star, consisting of pure absorption lines. This is a signature of a smaller emitting region as the equatorial disc is fully dissipated. However, as the equatorial disc forms and expands, the source enters the Be phase and the Balmer (and sometimes  \heI) emission lines start to appear. These emission lines are the result of recombination radiation from ionized gas in the equatorial disc around the Be star.

We have estimated the stellar parameters by comparing the observed spectrum to a grid of synthetic spectra. The medium resolution spectra taken on the night of 2012 Oct 27 were used as its H$\alpha$ line is the least affected by emissions from the Be circumstellar disc. We used the BSTAR2006 grid of metal line-blanketed, non-LTE, plane-parallel, hydrostatic model atmospheres of \citet{Lanz2007}, which was generated with the code TLUSTY \citep{Hubeny1995}. The synthetic spectra were calculated using the SYNSPEC program (version 49) for a range of effective temperatures between 15000 and 30000 K with a step of 1000 K and surface gravity 1.75$\le$\logg$\le$4.75 with a step of 0.25. We adopted a model atmosphere with half-solar chemical composition (i.e. LMC metallicity) and adopted a microturbulent velocity of $\xi$ = 5 \kms which is appropriate for our spectral type. The model spectra were also convolved with a Gaussian with FWHM~=~4.2 \AA\ to match the resolution of our spectra and broadened to \vsini = 285 \kms(see Section~\ref{rotbroad} for derivation of \vsini). Each spectrum from the grid is compared to the observed spectrum and the reduced \MyChi$^2$ were computed to test the goodness of fit. The best fit model parameter values estimated corresponds to \teff = 25000 K and \logg = 3.5, values that are consistent with the spectral classification of \asource as a B1 giant star. Fig.~\ref{plot_broad} (bottom panel) shows the observed combined spectrum taken on the night of 2012 Oct 27 compared to the best-fit model.

\subsection{Line profile evolution}

\subsubsection{The blue spectra}

\begin{figure}
\scalebox{0.5}{\includegraphics{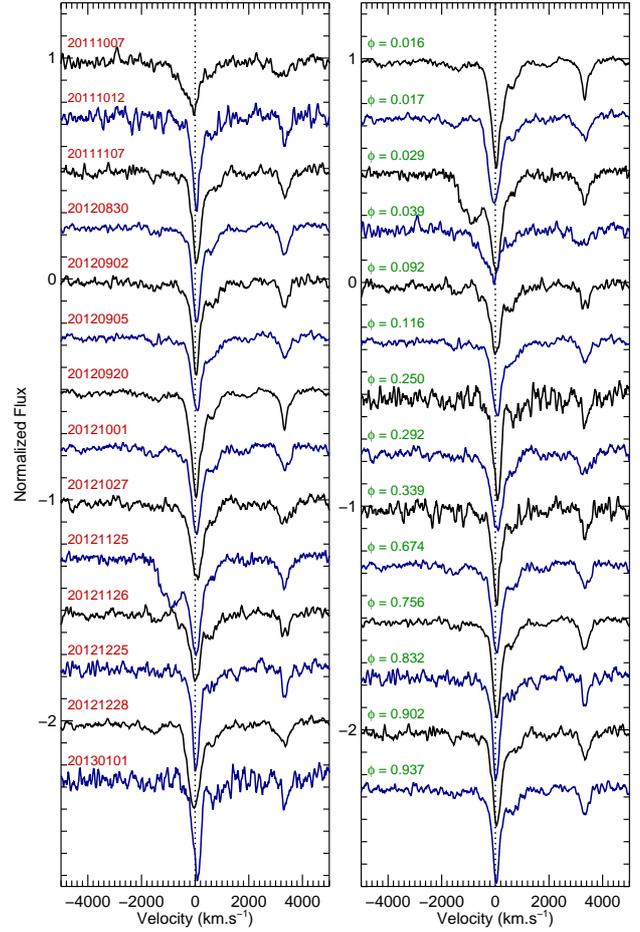}}
\caption{Evolution of the H$\gamma$ line profile sorted in time sequence ({\it left}) and by orbital phase ({\it right}). The dotted lines indicate the rest velocity.}
\label{plot_hgamma} 
\end{figure}

\begin{figure}
\scalebox{0.5}{\includegraphics{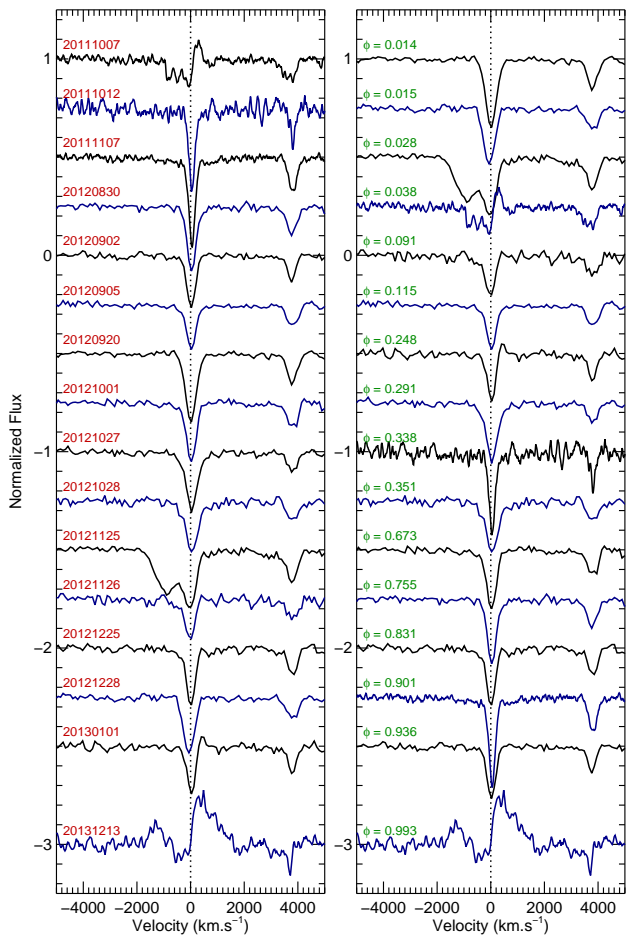}}
\caption{Evolution of the H$\beta$ line profile sorted in time sequence ({\it left}) and by orbital phase ({\it right}). The dotted lines indicate the rest velocity.}
\label{plot_hbeta} 
\end{figure}

The evolution of the H$\gamma$ and H$\beta$ line profiles in time and orbital phase are shown in Fig.~\ref{plot_hgamma} and \ref{plot_hbeta} respectively. The emission line profiles clearly vary in shape and strength over the orbital cycle. Outside periastron ($\phi > $ 0.1) the H$\gamma$ and H$\beta$ line profiles are seen in pure absorption except for the latest spectra taken on 2013 Jan 01 whose H$\beta$ shows evidence of a weak redshifted emission (P--Cygni profile). The strength and width of these Balmer lines vary with the optical brightness, they are narrower and deeper when the source fades (the equatorial disc is larger). For comparison, see the H$\gamma$ profile from the spectra taken on 2012 Oct 27 (near optical maximum) and spectra taken on 2012 Dec 25 (at outburst) and 2013 Jan 01 (the source fades).

During orbital outburst ( $\phi <$ 0.1) the source displays enhanced activity and its Balmer lines are much broader compared to those taken outside outburst in the same orbital cycle. The spectrum taken on 2012 Nov 25 ($\phi$ = 0.029) displays a strong blue-shifted absorption component with a velocity of $\sim$ 850 \kms in both the Balmer and \heI~ lines. One day later, the blue-shifted absorption has completely disappeared. The H$\beta$ line of the 2011 Oct 07 ($\phi$ = 0.038) spectrum exhibits a P--Cygni profile with three absorption components at 815, 504 and 34 \kms.

\begin{figure}
\scalebox{0.5}{\includegraphics{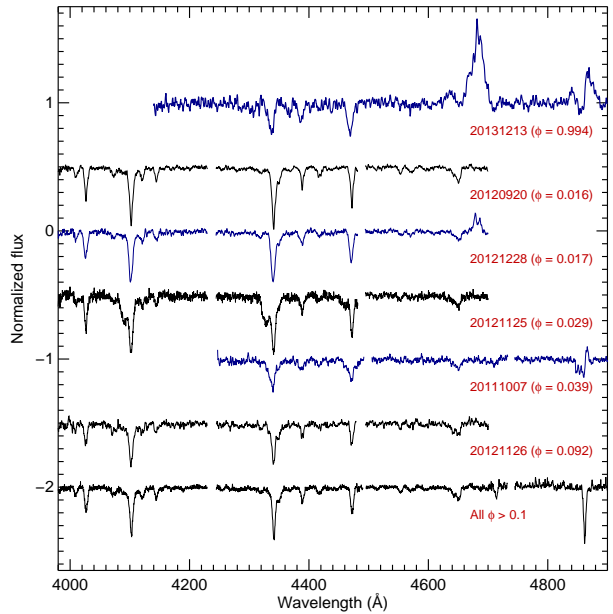}}
\caption{SAAO 1.9 m and SALT blue spectra of \asource taken around periastron and the average of all SALT spectral taken outside periastron (bottom). The gaps in the spectra are CCD gaps.}
\label{plot_highres} 
\end{figure}

Fig.~\ref{plot_highres} shows the SAAO 1.9~m and SALT blue spectra taken during outburst that are arranged according to the orbital phase, and the average spectra of all spectra taken outside the outburst (bottom spectrum). We have detected the presence of the symmetric double-peaked \heII~$\lambda$4686 emission line with a peak-to-peak separation of 510~\kms which is only present in the spectra taken at orbital phase 0.95$<\phi<$0.02 and during optical low state of the source. At optical low state the Be equatorial disc is well developed and dense enough for the NS to accrete at periastron. The \heII~emission line detected in the SALT spectra obtained on 2012 Dec 28 has a double-peaked profile with equivalent width (EW) 1.35$\pm$0.04 \AA. The appearance of this line was accompanied by brightening of $\sim$0.5 mag in our $I$-mag OGLE~IV data. The spectra taken with the SAAO 1.9~m on 2013 Dec 13 ($\phi$=0.994, top spectrum in Fig.~\ref{plot_highres}) also shows a very strong (much stronger than H$\beta$) and very broad \heII~emission line with an EW of 15.17$\pm$0.34 \AA. The features to the blue and red of \heII~$\lambda$4686 in our spectrum are the (\CIII + \OII) blend $\lambda$4640--50 \AA~and \heI~$\lambda$4713 \AA, respectively. Unfortunately, this optical outburst falls in the gap of the OGLE~IV data, therefore we could not estimate its amplitude. However, the observed outburst just before it has an amplitude of $\Delta I \sim$ 0.3 mag, so we believe an increase in brightness of $\sim$ 0.3 mag or slightly higher took place during our SAAO 1.9~m observation. We note that there is no sign of any \heII~emission in the spectra taken with SAAO 1. 9 m on the following night (2013 Dec 14).

\subsubsection{The \halpha line}

\begin{figure}
\scalebox{0.5}{\includegraphics{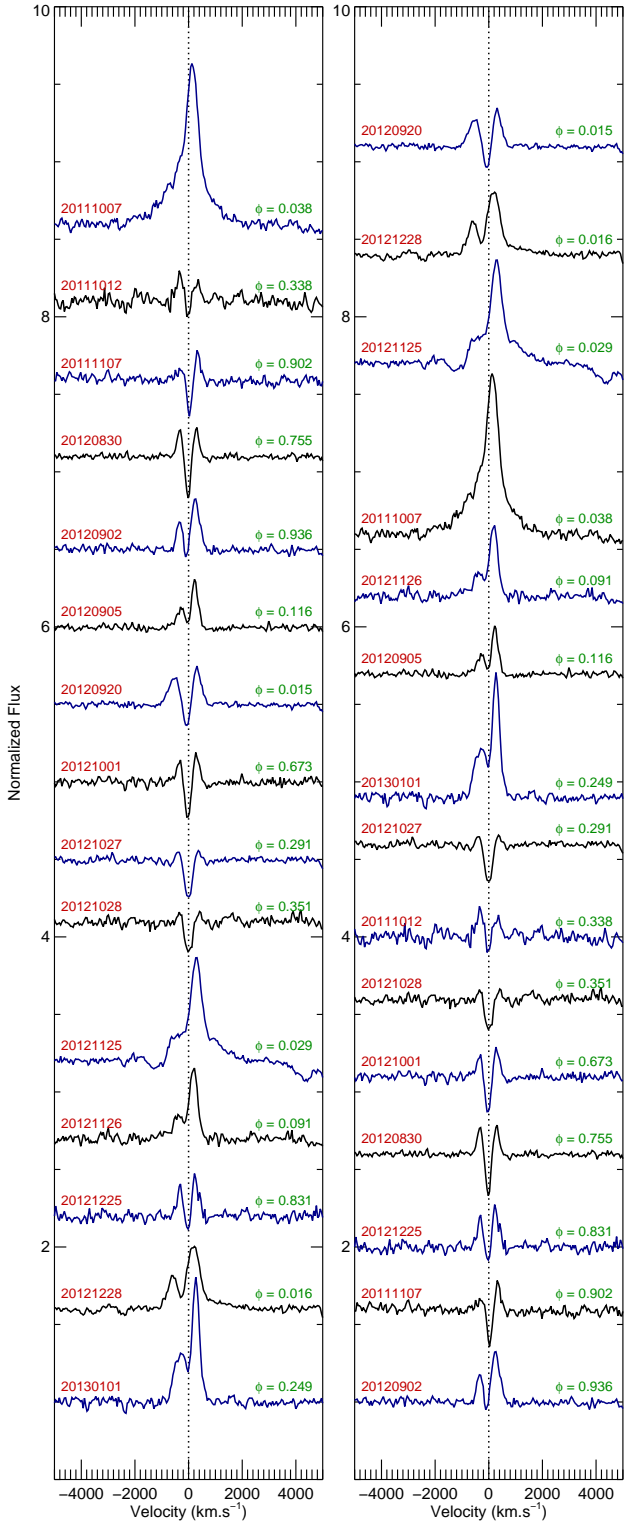}}
\caption{Evolution of the H$\alpha$ line profile sorted in time sequence ({\it left}) and by orbital phase ({\it right}). The dotted lines indicates the rest velocity. Note the change in the line profiles from double-peaked to singled-peak at periastron ($\phi$ = 0.038) and symmetric shell profile far from periastron.}
\label{plot_halpha} 
\end{figure}

\begin{figure}
\scalebox{0.5}{\includegraphics{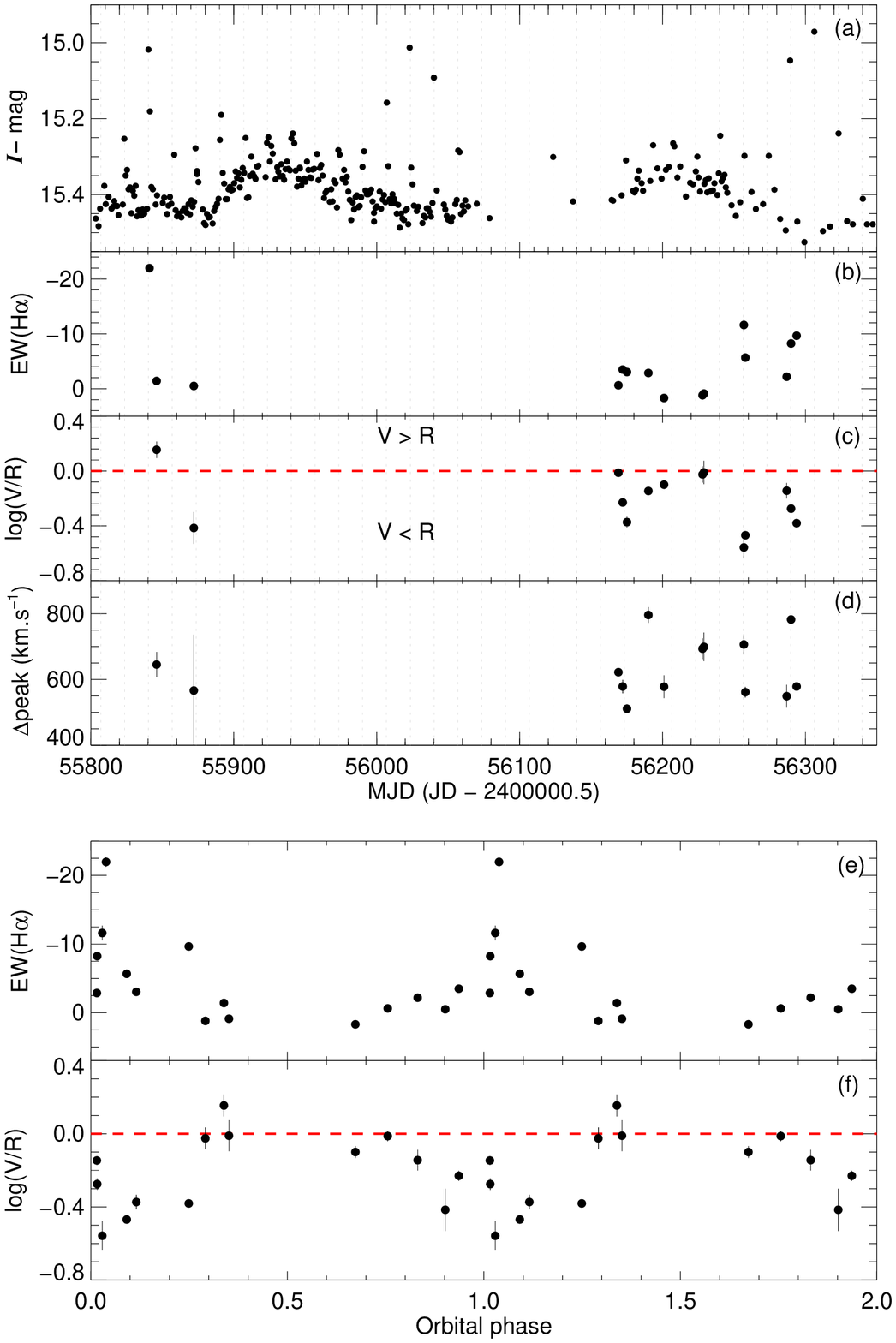}}
\caption{\halpha Long-term evolution of the (a) brightness ($I$ mag), (b) EW(\halpha), (c) logarithm of the ratio V/R and (d) separation of the peaks of the \halpha profile. (e) and (f): Orbital variability of the EW(\halpha) and log(V/R), respectively.}
\label{plot_fithalpha} 
\end{figure}

The \halpha emission line in Be stars is known to show a wide range of variability, over both short and long temporal scales. This is linked to the variation in the structure of the circumstellar disc which forms and depletes on timescales of years to decades \citep{Reig2005, Reig2011}. One can trace the evolution of such a disc by monitoring variations of \halpha emission line strength, whose flux is dominated by the emission from the equatorial disc of the Be star. During our observations, we can see both short and long term variations in the \halpha profile.  The short-term variation is related to orbital motion of the NS which interacts with the disc around the Be star near periastron, whereas the longer term variation is related to variation in the size of the Be equatorial disc that forms and depletes on a timescale of years in a quasi-periodic manner. Fig.~\ref{plot_halpha} shows the evolution of the \halpha  emission from \asource, in which the line profile is highly variable, both in strength and shape. It exhibits a wide range of profiles such as single-peaked emission (2011 Oct 07), double-peaked emission (e.g. 2013 Jan 01) and shell profile in which the central absorption that separates the two emissions extends below the stellar continuum (e.g. 2012 Aug 30).

The singled-peaked profile is only seen in the spectra taken very near to periastron ($\phi\sim0.04$) on 2011 Oct 07, during which the EW is at its maximum value (EW(\halpha) $\sim$ 22 \AA). The shell profile shows both symmetric and asymmetric profiles. Furthermore, the symmetric shell profiles are seen in the observations taken far from periastron and when the \halpha EW is lower (smaller equatorial disc). The double-peaked emission profile is highly asymmetric and also shows a variation in the relative strength of its violet (V) and red (R) components, with the red component much stronger than the blue components (V < R) for most of our observations. Only the earliest spectra taken on 2011 Nov 12 show a marginally blue-dominated peak.

The spectrum taken during the orbital outburst on 2011 Oct 07 shows an enhanced \halpha activity with a red-shifted single peaked emission superimposed on a broad emission wing extending to velocities of more than 2200 \kms in both blue and red.  The \halpha profile in the spectra taken on 2012 Nov 25 and 26 ($\phi$ = 0.029 and 0.091, respectively) also shows these broad emission wings with a blue-shifted absorption component, similar to the other Balmer lines.

We measured the line parameters by fitting the normalized spectrum with two or three Gaussians depending on whether the H$\alpha$ line exhibits single-peaked with broad wing, double-peaked or shell profiles. The third Gaussian component (in absorption) is needed for the shell profile as the central absorption line goes below the continuum. Fig.~\ref{plot_fithalpha} shows the evolution of the brightness, EW(\halpha), log(V/R) and the separation of the V and R peaks of the \halpha profile. We also plot the orbital variability of the EW(\halpha) and the ratio V/R using the ephemeris derived from the optical outburst above. The EW(\halpha) and the ratio V/R are obviously modulated by the orbital phase. Moreover, these two parameters show clear anti-correlation with each other around periastron (Fig.~\ref{plot_fithalpha}).

\subsection{Spectral classification}

\begin{figure*}
\scalebox{0.5}{\includegraphics{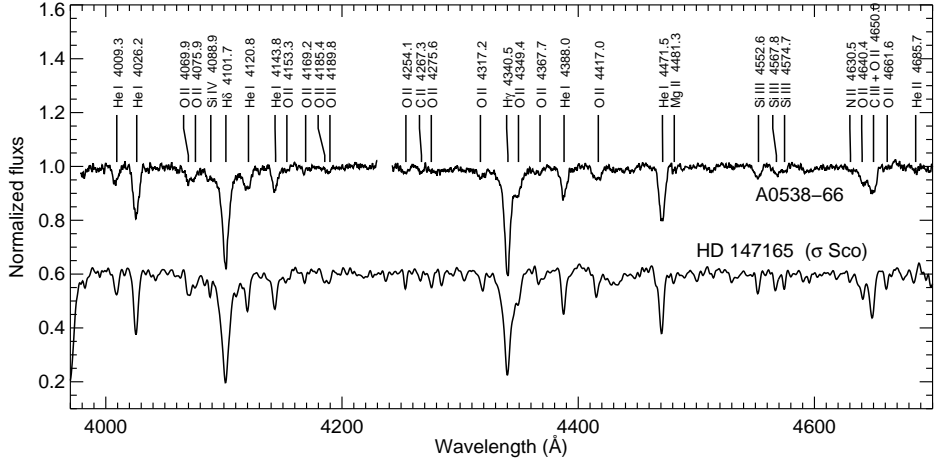}}
\caption{Average blue spectra of \asource in which all identified lines are marked. The spectrum of the B1~III star $\sigma$ Sco (HD 147165) from \citet{Walborn1990} is also displayed for comparison. The gaps in the spectra are CCD gaps.}
\label{classification} 
\end{figure*}

We rectify all spectra by fitting a low-order spline to the continuum and then we combine those taken outside periastron. The resulting average high resolution blue spectrum ($\lambda\lambda$4000--4700) is shown in Fig.~\ref{classification}, in which all main spectral features have been identified. The spectrum is mostly dominated by Balmer and neutral helium (\heI) lines and does not show any evidence of the \heII~absorption lines (\lambdaf4200, \lambdaf 4541, \lambdaf 4686) above the noise level, indicating a spectral type later than B0.5. The weakness of the \MgII $\lambda$4481 line suggests a spectral type earlier than B2. The strength of the absorption lines of the \CIII + \OII blends at 4650~\AA~ also indicates a B1 spectral-type source. The ratio \SiIII \lambdaf4552/ \SiIV \lambdaf4089 confirms the classification of the optical counterpart as a B1 star.

The luminosity class was determined using the ratio between the \SiIII $\lambda$4552 and the \heI $\lambda$4387 lines, which suggests a luminosity III (giant) star. The strength of the \OII lines especially at 4349, 4070 (blend), 4417 and 4640 (blend) also indicates a luminosity class of III. The strength of these \OII lines increases with luminosity. We conclude, therefore, that the optical counterpart of \asource is a B1e~III star, which is slightly earlier than the previously reported spectral type (B2~III--V). As a comparison the spectrum of a standard B1~III (HD 147165) from the \citet{Walborn1990}  spectral atlas is also plotted in Fig.~\ref{classification}.

The observed brightness of \asource in the absence of the equatorial disc is $V$ = 14.885 (see section~\ref{discless_phase}). Assuming the LMC distance modulus to be 18.5 \citep{Alves2004} and an extinction value of $A_{\rm V}$ = 0.38 \citep{Imara2007}, gives an absolute magnitude of $M_{\rm V}$ = -- 4.0 which indicates a spectral type in the range B1~III--IV \citep{Humphreys1984, Gray2009}, consistent with the spectral type derived spectroscopically.

\subsection{Orbital solution}

\begin{figure}
\scalebox{0.5}{\includegraphics{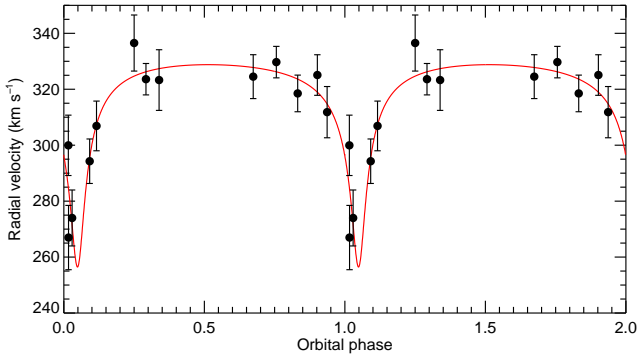}}
\caption{Radial velocity curve of the Helium lines for \asource. The red line shows the best-fit radial velocity.}
\label{plot_radvel} 
\end{figure}

The radial velocities have been measured by cross-correlating each individual spectrum with a rotationally and instrumentally broadened BSTAR2006 synthetic B1~III (\teff~=~25000 K and \logg~=~3.5) template spectrum. We have only used the neutral helium lines (\heI~$\lambda$4026, $\lambda$4143, $\lambda$4387, $\lambda$4471) as they are less contaminated by any emission from the disc compared to the Balmer lines. We did not include the spectra taken on 2011 Oct 07 in our analysis because of its low S/N, few \heI~lines and the presence of an H$\beta$ P--Cygni profile. Fig.~\ref{plot_radvel} shows the resulting radial velocity curve folded on the orbital ephemeris above.

\begin{figure}
\scalebox{0.5}{\includegraphics{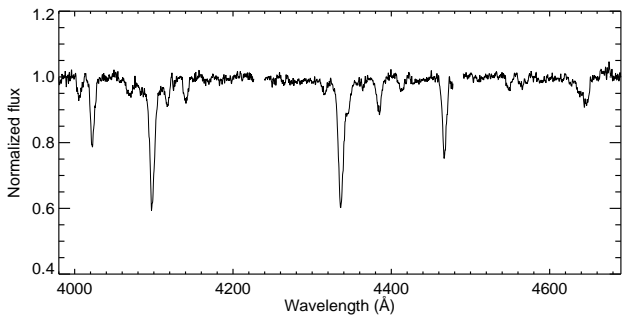}}
\caption{Average of velocity-corrected spectra taken around periastron.  The gaps in the spectra are CCD gaps.}
\label{plot_periastron} 
\end{figure}

We have attempted to fit this radial velocity curve with a Keplerian orbit by fixing the orbital period to $P_{\rm orb}$=16.6409d. The orbital solution from the fit is overplotted in Fig.~\ref{plot_radvel} and the best fitting parameters are shown in Table~\ref{rv_fitres}. In Fig~\ref{plot_periastron}, we have plotted the average velocity-corrected spectrum of spectra taken around periastron in which the cores of the \heI lines appear undistorted, and hence suffered minimal emission.

\begin{table}
\centering
\caption[]{Orbital parameters of \asource.}
\label{rv_fitres}
\begin{tabular}{lc}
\hline
\hline
Parameter                   &           \asource         	        \\
\hline
$P_{\rm orb}$ (d)        &       16.6409 (fixed)        \\
$T_0$ (HJD-2,450,000.5)       &     55673.98 (fixed)      \\
$e$                         &	    0.72~$\pm$~0.14	 	  \\
$\omega$ (deg)              &	    183.0~$\pm$~9.4	          \\
$\gamma$ (km s$^{-1}$)      &	    318.9~$\pm$~2.7	  \\
$\phi_{\rm peri}$           &	    0.04~$\pm$~0.01	   \\
$K_{\rm opt}$ (km s$^{-1}$) &        35.39~$\pm$~16.28	   \\
$a_1 \sin i$ (R$_{\odot}$)  &	    8.07~$\pm$~3.71  \\
$f(M)$ (M$_{\odot}$)        &    0.025~$\pm$~0.019     \\
\hline
\end{tabular}
\end{table}

\subsection{Rotational broadening}
\label{rotbroad}

Be stars are believed to rotate close to break-up velocity \citep{Townsend2004}.  The projected rotational velocity was estimated by using the relations between the rotational velocity and the full width half maximum (FWHM) of  \heI lines  given in \citet{Steele1999}. We measured the FWHMs of four  \heI lines ($\lambda$4026, $\lambda$4143, $\lambda$4387, $\lambda$4471) from the 2012 Oct 27 spectrum which is the least affected by contamination by emission from the equatorial disc. The measured FWHMs were then translated to a rotational velocity and its weighted average was estimated. After correcting for instrumental broadening by subtracting in quadrature the instrumental resolution of 50 \kms, we obtained a weighted average of \vsini~= 285 $\pm$ 27 \kms.

Another way of estimating the rotational velocity is to compare the observed spectra with a grid of synthetic spectra broadened with different values of the velocity. We used the same model spectrum as above (BSTAR2006 grid) and adopted a model atmosphere with half-solar chemical composition, \teff~= 25000 K, \logg = 3.5 and microturbulent velocity of 5 \kms. We convolved this synthetic spectrum with a set of rotational profiles with velocity from 100 -- 500~\kms in steps of 5 \kms, using the ROTIN3 routine. We chose a linear limb-darkening coefficient of 0.37 which is appropriate for our spectral type \citep{Reeve2016}. The derived projected rotational velocity which minimizes \MyChi$^2$ corresponds to \vsini = 290 \kms, consistent with the value derived from the FWHMs of the \heI lines .

\section{Discussion}

\subsection{Long-term optical variations}

BeX systems are known to exhibit large amplitude long-term photometric variability on timescale of years to decades. Such quasi-periodic superorbital variations have previously been reported in a number of BeX systems \citep[see][]{McGowan2008, Rajoelimanana2011}. \citet{Alcock2001} suggested that the 420.8 d superorbital modulations in the first 5 years of MACHO optical light curves of \asource are related to the variation in the size of the Be equatorial disc. As described in section~\ref{optical_lightcurve}, this superorbital variation appeared to have shortened in the second half of the MACHO. Hovever, in spite of a significantly reduced amplitude, the periodogram of the recent OGLE~IV data still shows a highest peak at 423.74d, very close to that previously reported, as well as additional power around 200d. These shorter period peaks are due to the change in the overall shape of the superorbital modulation in the OGLE~IV light curve compared to the MACHO era.

\subsubsection{Disc-less phase (optical high state)}
\label{discless_phase}

The circumstellar disc is the main reservoir of material available for accretion by the NS. We would not, therefore, expect to see any outburst when the disc is absent or very weak (i.e. during the disc-less phase). The MACHO $V$ and $R$ light curves show flat optical maxima that last $\sim$ 200 d, during which the colour is bluest and does not change. This suggests that the system has entered its disc-less, normal B star phase. We therefore infer that the brightness ($V$ = 14.885, $R$ = 15.025) and colour ($V - R$ = --0.14) measured at those times correspond to the true intrinsic magnitudes and colour of the underlying B star. This was confirmed by our optical maximum spectra which are dominated by the photospheric spectrum of the underlying B star. The optical high state in the OGLE~IV observations occurred near 2012 Oct 27 and the observed \halpha line profile at that time appeared to be in absorption. However, some weak orbital outbursts are still seen in the light curve which indicates that the Be equatorial disc is not dissipating completely, and the NS can still influence it. This residual disc material at the optical maxima in the OGLE~IV data still partially obscures part of the B star, resulting to a decrease in the overall amplitude of the 420.8d compared to the disc-less phase during the MACHO observations.

\subsubsection{Be phase (optical low state)}
\label{bephase}

We are viewing the Be equatorial disc in \asource nearly edge-on (see section~\ref{geometry}) which means when it forms it will mask part of the hotter (brighter) B star, causing the optical brightness to decline. The presence of the circumstellar disc makes the system redder and gives rise to Balmer and sometimes \heI emission lines. The passage of the NS near periastron has an affect on the structure of the Be circumstellar disc and its emission properties. If the disc is large enough, then the NS interacts and accretes from it near periastron which gives rise to periodic outbursts seen in X-ray and optical. During the orbital outburst the colour of the system reddens slightly, which suggests that the additional brightening originates from the cooler equatorial disc.

The strength of outburst also varies significantly throughout the superorbital cycle, which means that the observed effect of the periastron passage of the NS depends strongly on the density and extent of the equatorial disc. The outbursts are much stronger when the disc is denser and larger (at optical minima) and do not occur when the disc is very weak or absent (disc-less phase).

Furthermore, the light curve of \asource folded on the period of 16.6409~d shows a double peak profile with two peaks occurring at phase~0.0 and 0.071 (see Fig.~\ref{fold_all}). This suggests that there is a misalignment between the spin axis of the Be equatorial disc and the orbital plane of the NS. A spin-orbit misalignment has previously been seen in some BeX systems \citep[e.g. SXP327, ][]{Coe2008}. The phase of the two optical peaks ($\phi$=0.0 and 0.071) as well as the time of periastron passage ($\phi$=0.04) were plotted in the Fig.~\ref{plot_geom}. The presence of two peaks in \asource has also been reported by \citet{Corbet1997} in which they suggest that the increase in relative velocity near periastron may cause the dip in brightness between the peaks. We note that they observed the dip on 1997 Feb 02 which corresponds to phase~0.05 (very near periastron) in our new ephemeris, and so we interpret this phenomenon as two superposed peaks resulting from the misaligned equatorial disc.

The appearance of the \heII~\lambdaf 4686 emission line in our spectra indicates the presence of hard ionizing radiation with energies above 54 eV which means \asource was in an X-ray "on" state at that time. The observed \heII~\lambdaf 4686 emission line has a double-peaked profile and is probably due to radiative recombination in a photoionized accretion disc around the NS rotating with Keplerian velocity. The \heII~emission line appears only in the spectrum taken near periastron and is strongest at phase~0 (optical outburst, EW$\sim$15.17 \AA~on 2013 Dec 13) where the NS is predicted to interact and accrete from the circumstellar disc. This suggests that the accretion disc in these systems is transient and is only present around periastron. The \heII~emission line has been previously detected in \asource, all of which occurred near phase~0 \citep[e.g.][]{Charles1983, Corbet1985}.

\subsection{\halpha emission}

The \halpha emission line in BeX systems originates from the geometrically thin, nearly Keplerian circumstellar disc around the Be star \citep{Porter2003}. Therefore, studying the evolution of its profiles will allow us to estimate the long-term structural changes that occur within the circumstellar disc. The values of EW (\halpha) show considerable variability throughout our SALT observations. Because of the short orbital period and large eccentricity in \asource, the periastron passage of the NS has a significant effect on the evolution of the Be circumstellar disc. The periodic gravitational influence of the NS near periastron in such a narrow orbit prevents the Be circumstellar disc from growing larger and causes truncation \citep{Reig1997, Okazaki2001}. 

The \halpha emission profile in \asource is a strong function of orbital phase. Far from periastron, it shows a symmetric shell profile, characteristic of an unperturbed equatorial disc viewed nearly edge-on and rotating with Keplerian velocity distribution. However, near periastron the line profile is double-peaked and highly asymmetric, becoming very strong and single-peaked at periastron. Near periastron, the circumstellar disc may be distorted by the NS' gravity, and this will make it highly non-axisymmetric with an enhanced density toward the compact object. Based on our orbital solution of $\omega$=183$\degr$ and $e$ = 0.72, the NS (and hence the high density region of the equatorial disc) is moving away from us near periastron (see Fig.~\ref{plot_geom}). Therefore, we would expect an increase in the strength of the red peak of the \halpha profile at phase $\sim$ 0.0--0.1 (on the right of the Be star in Fig.~\ref{plot_geom}). However once the NS is far from periastron the \halpha line profile will return to a symmetric shell profile again. This is what we have seen in our spectra taken on 2012 Dec 25.

The spectra obtained at periastron (see $\phi$=0.038 and 0.029) are single-peaked and also have a broad wing component. This suggests that at periastron we are viewing the circumstellar disc at a lower inclination angle than it was before (outside periastron). This change in the disc inclination implies that at periastron the tidal torque from a misaligned companion neutron star tends to warp the equatorial disc towards the binary orbital plane which has a lower inclination angle than the disc. However, as a result of warping, the inner region close to the Be star, where the velocities are largest, can be at a still higher inclination angle. This high velocity region contributes to the observed broad wings in the profile. Similar behaviour has been observed in 4U 0115+63/V635 \citep{Negueruela2001}. \citet{Martin2011} also reported that the tidal torque from the NS could warp the equatorial disc in BeX system with short orbital periods if the orbit is eccentric and misaligned with the equatorial disc.

\subsection{Geometry of the system}
\label{geometry}

The behaviour of the brightness and colour of \asource (brighter when bluer) indicates that the inclination angle of the circumstellar disc is large ($i \geq90\degr-\alpha $, where $\alpha \leq 10\degr$ is the opening angle of the disc). We note that the majority of BeX systems show the opposite behaviour, where the source is redder when brighter \citep[see][]{Rajoelimanana2011}. In addition, the presence of the shell profile in the \halpha line confirms that we are viewing the Be equatorial disc nearly edge-on \citep{Hummel2000}.

Using our values for the projected rotational velocity (\vsini~= 285 \kms) and measured peak separation, and assuming a Keplerian rotation of the disc, the outer radius of the \halpha emitting region can be estimated using the equation \citep{Huang1972}:

 \begin{equation}
     \left( \frac {R_{\rm out}}{R_{1}}\;\right)
       = \;       \left( \frac{2\,v\,\sin{i}}
                {\Delta V} \;\right)^{2}  ,
  \label{Huang1972}
  \end{equation}

\noindent
where $R_{1}$ is the radius of the Be star and $\Delta V$ is the peak separation.  

\begin{figure}
\scalebox{0.5}{\includegraphics{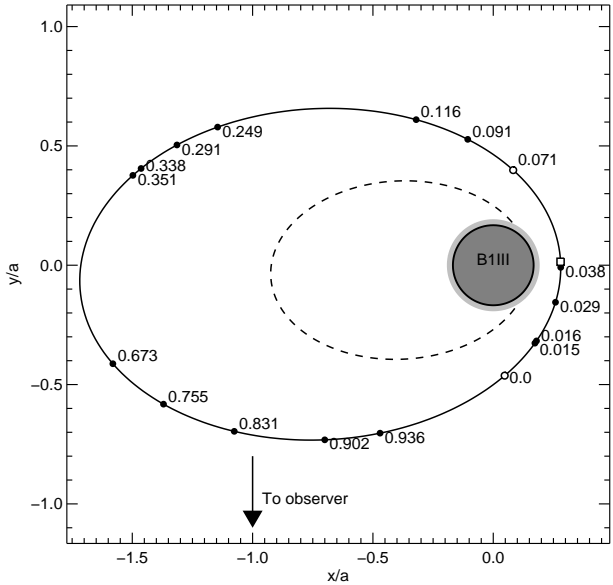}}
\caption{Geometry of the system, showing the relative orbit of the NS around the Be star computed using our orbital solution. The coordinates are in units of the semimajor axis. phase~0 is defined as the time when the outburst reaches its maximum brightness. The phase of the periastron (square), the two optical outburst peaks (circle) and the SALT observations (filled circle) were marked. The black and grey circles represents the optical Be star with radius R$_1$ =10.0 \rsun and the equatorial disc with radius $R_{\rm out}$= 1.15 $R_{1}$, respectively. The dashed ellipse represents the Roche lobe radius at each orbital phase for masses of $M_{1}$ = 8.84 \msun and $M_X$ = 1.44 \msun. The NS is moving counter-clockwise around the Be star. We note that the figure is in scale.}
\label{plot_geom} 
\end{figure}

The presence of the \heII~$\lambda$4686 line in the spectra taken at $\phi$ = 0.017 (2012 Dec 28) at the end of our SALT campaign suggests that mass transfer onto the NS is again occurring in the system. To estimate the outer radius around that period of time, we measured a peak separation of 540 \kms from spectra taken far from periastron and at the same orbital cycle (2012 Dec 25). This gives an outer radius of  $R_{\rm out}=1.15~R_{1}$ (shown as the grey circle in Fig.~\ref{plot_geom}). We note that we could still see weak outbursts in the optical high state during the OGLE observations, therefore the value of our \vsini  and hence the outer radius of the \halpha emitting region that we measured is slightly underestimated.

\begin{figure}
\scalebox{0.5}{\includegraphics{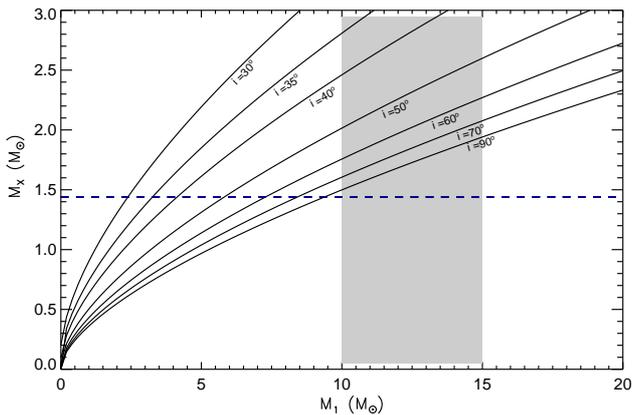}}
\caption{Mass constraints for the two components of the binary system. The lines have been calculated using our mass function for the labelled inclination values. The range of possible masses for the B1~III optical star (10.0 \msun < $M_1$ < 15.0 \msun) are represented by the grey region. The dashed horizontal line represents the standard mass of the neutron star ($M_{X}$=1.44\msun).}
\label{mass_range} 
\end{figure}

Our spectroscopic study is essentially ``single-lined" as, until \asource is recovered as an X-ray pulsar, the only kinematic constraints on the NS are provided by \heII~$\lambda$4686, which we assume to be emitted close to the accretion disc, and that only appears near periastron passage.  Our radial velocity curve therefore provides only the mass function. Furthermore, the binary inclination, which clearly must be different from the inclination of the Be equatorial disc, is also unknown. In Fig.~\ref{mass_range} , we have used our mass function to plot the range of possible masses of the Be companion star and the NS for several values of the binary inclination. If we consider a neutron star mass range of 1.44 \msun $\le M_{\rm X} \le$ 3.0 \msun~and a corresponding range for the Be star mass of  10.0 \msun  $\le M_{\rm 1} \le$  15.0 \msun, then clearly we require $i_{\rm orb}$ > $34\degr$ (see Fig.~\ref{mass_range}).

However, there are additional constraints. The absence of X-ray eclipses in the light curves of \asource \citep[see][]{Skinner1980a} implies that $i_{\rm orb} \le 75\degr$ for a B1~III star with radius $R_1$ = 10.0 \rsun. This is in good agreement with our interpretation that the double peaked outbursts are related to the disc-orbit misalignment in the system. Using the mass function, an upper limit of $i_{\rm orb} \le 75\degr$, and adopting a canonical NS mass of 1.44 \msun, we estimate an upper limit of 8.84 \msun~for the mass of the Be donor. Therefore, unless the NS is heavily over-massive, the Be star is under-massive for its spectral type of B1~III.  However, this is highly typical of high mass X-ray binaries in general, where the evolutionary path has led to a donor that has significantly different properties of those of a normal main-sequence equivalent \citep[see e.g.][]{Rappaport1983}.

The radius of the Roche lobe can be computed using \citet{Eggleton1983} formula 

\begin{equation} \label{Roche_radius}
\frac{R_L}{A} = \frac{0.49 q^{2/3}}{0.6q^{2/3} + \ln (1+q^{1/3}) } ~~~~~,~~~~~~(0<q<\infty)
\end{equation}
\noindent
where q = M$_1$/M$_X$ is the mass ratio and $A$ is the orbital separation. The Roche lobe radius as a function of the orbital phase using q = 8.84/1.44~$\sim$ 5.86 is plotted in Fig.~\ref{plot_geom}. We note that corotation is only satisfied close to periastron, and so our calculation for the rest of the orbit is only very approximate. With these parameters, the Be star (with radius $R_1$ = 10.0 \rsun) would fill its Roche lobe at phase~0.025 ($\sim$0.5 day before periastron) and would overfill it at periastron ($R_{\rm Lper}$ = 8.98 \rsun). However, the absence of optical and X-ray activities during the extended optical maxima indicates that the naked B star may not fill its Roche lobe even at periastron, but this requires larger masses for both components. For inclinations $i_{\rm orb} \le$ 75\degr, we derived a lower mass limit of 1.73 \msun for the NS and 11.8 \msun for the donor star in order to get them to be within the Roche lobe.

\section{Summary}

We have presented an analysis of the long-term behaviour of the BeX source \asource using the extensive light curves from the MACHO and OGLE~IV projects and more recent high S/N optical spectroscopy from SALT. The $\sim$420d superorbital modulations reported by \citet{Alcock2001} have shortened in the second half of the MACHO data but are still seen in the OGLE-IV data with a period of $\sim$424d. This indicates that the superorbital modulation persists over timescales of decades. Using extensive timebase of the combined MACHO and OGLE~IV datasets, we have refined the orbital period to 16.6409$\pm$0.0003d. 

Both broad-band and high-resolution SALT spectra confirmed the classification of A0538-66 as a B1~III star. We have derived the orbital parameters of the binary system, obtaining an eccentricity of  $e$=0.72$\pm$0.14, $K_{\rm opt}$=35.39$\pm$16.28 \kms and $\omega$=183.0$\pm$9.4$\degr$. The resulting mass function suggests the mass donor Be star is significantly undermassive for its spectral type unless the NS mass far exceeds the canonical 1.44 \msun. We favour a mass $M_X\ge$ 1.73~\msun for the NS in \asource. The shape of the \halpha emission line profile strongly depends on the binary orbital phase. It shows a symmetric shell profile when far from periastron, becoming double-peaked near periastron and then single-peaked at periastron. We interpret these changes as due to the tidal torque from a misaligned neutron star on the Be equatorial disc during periastron passage.

\section*{Acknowledgments}

Some of the observations reported in this paper were obtained with the Southern African Large Telescope (SALT) under program 2011-3-RSA$\_$UKSC-008 (PI: Charles), 2012-1-RSA$\_$UKSC-005 and 2012-2-RSA$\_$UKSC-003 (PI: Rajoelimanana). The OGLE project has received funding from the National Science Centre, Poland, grant MAESTRO 2014/14/A/ST9/00121 to AU. This paper utilizes public domain data obtained by the MACHO Project, jointly funded by the US Department of Energy through the University of California, Lawrence Livermore National Laboratory under contract No. W-7405-Eng-48, by the National Science Foundation through the Center for Particle Astrophysics of the University of California under cooperative agreement AST-8809616, and by the Mount Stromlo and Siding Spring Observatory, part of the Australian National University.

\bibliographystyle{mn2e}

\label{lastpage}
	
\end{document}